\documentclass{article}

\usepackage{amssymb}

\usepackage{amsmath}

\usepackage{graphicx}

\def\GeV{\nobreak\,\mbox{GeV}}

\begin{document}

\title{Exact treatment of  dispersion relations in  pp and
 p\=p elastic scattering}

 \author{\ \\
E. Ferreira\thanks{Email: erasmo@if.ufrj.br},\hspace{20pt} A. K. Kohara\thanks{Email: kendi@if.ufrj.br}, \\  \  \\
{\em Instituto de F\'{\i}sica, Universidade Federal do Rio de Janeiro,} \\
{\em 21941-972, Rio de Janeiro, Brasil}\\ \ \\
and \\ \ \\
J. Sesma\thanks{Email: javier@unizar.es} \\ \ \\
{\em Departamento de F\'{\i}sica Te\'{o}rica, Facultad de Ciencias,} \\
{\em 50009, Zaragoza, Spain}}

\maketitle


PACS{\hspace{10pt} 13.85.Dz; \and 13.85.Lg; \and 13.85.-t}


\begin{abstract}
Based on a study of the properties of the Lerch's transcendent,
exact closed  forms of  dispersion relations for amplitudes and for derivatives of amplitudes
 in pp and p\=p  scattering  are  introduced.
Exact and complete expressions are written for the    real parts and for their derivatives at $t=0$
based on  given inputs
for the energy dependence of the total cross sections and  of the slopes of the
imaginary parts.
The results are prepared for application in the analysis of forward scattering data
of the  pp and p\=p systems at all energies, where exact and precise
representations can be written.
\end{abstract}

\section{Introduction}

Elastic scattering in the pp and p\=p systems  is analytically very simple:  spin effects
neglected,  it is described by a single complex function of two variables ($s$, $t$).
This formal simplicity hides the fact that elastic scattering is a coherent nonperturbative
process  involving complicated dynamics. After four  decades of studies in the framework
of the modern theory of the strong interactions (QCD), still  no fundamental
microscopic model is successful  in the description of the imaginary
and real parts of the complex amplitude. In spite of the essentially  complex
fundamental dynamics, the experimental data and the observables on the differential elastic
cross sections show simple and regular   dependences on the energy $s$ and on the
transferred momentum $t$.   It seems that the global simple
behavior is actually a very consequence of the extreme internal complexity.

To build a bridge between data and  microscopic models, the differential cross sections
in the forward range can be represented in terms of a few parameters, with precision
and coherence. We present in this paper the
formulation, based on principles of Dispersion Relations (DR), that is appropriate for
the analysis of $d\sigma/dt$ data. Our treatment  uses two important developments.
One is the discovery of exact solution of the principal value (PV) integrals  that
occur in dispersion relations, based on our recent work on properties of the
mathematical function called Lerch's transcendent
 \cite{EF-K-Sesma3}. The second is the application of the
Dispersion Relations for Slopes (DRS) \cite{EF2007}, based  on the knowledge of the
$s$ dependence of the slopes of the imaginary parts of the amplitudes. The real
and imaginary parts of the amplitudes in the forward direction are treated as independent
functions, as required by quantum mechanics, with  connections  determined by the causality
and analyticity foundations of the dispersion relation. The usual assumption of equal real
and imaginary slopes is here forbidden.

In the best explored forward region the analysis of pp and p\=p scattering is
affected by the comparatively small magnitude of the real amplitude (usually expressed through the
parameter $\rho$ that gives the ratio of the real to the imaginary part at $t=0$), whose sign and
strength must be studied through interference with the Coulomb
interaction. The extraction of precise information on the real part is very difficult, with
consequences on the determination of the optical point in the imaginary part that leads to the
total cross sections. In these conditions, it is essential to use the full potentiality of
theoretical controls, such as DR and DRS.

The original forms  of Cauchy principal value integrals (PV) occurring in the DR
are not very practical in  calculations.  Local forms,
called  Derivative Dispersion Relations (DDR), are  more confortable, and have been used
in the analysis of  the data.  After a period in
which the knowledge of DDR was limited to approximations not valid
for low energies, the  connection between integral and local forms
has been put in exact terms \cite{Menon,EF-Sesma1,EF-Sesma2}, giving
mathematically correct relations between real and imaginary
parts of the complex amplitude. These local forms consist of double \cite{Menon}
and single   \cite{EF-Sesma1,EF-Sesma2} infinite series, of fast convergence
in the applications.

In the present work we introduce new results \cite{EF-K-Sesma3}
for the exact  DR forms, written in terms of the function called Lerch's transcendent.
From now on, terms of the input imaginary amplitudes used in pp and p\=p phenomenology
have their real counterparts  written in compact analytical expressions.
The new expressions for the exact forms of DR have their properties discussed
and are used to draw consequences of the input form of the imaginary amplitude
(namely of the total cross section) proposed by the Particle Data Group/Compete Collaboration
\cite{PDG} (PDG). Since we write exact forms, we call attention for the importance of
the influence of the subtraction constant $K$, that cannot be ignored at low energies.

We go one step further, and explore, again in exact terms, the idea of the
Dispersion Relations for Slopes (DRS) \cite{EF2007} that was introduced  in the year 2007
and shown to be important for the analysis of pp and p\=p
scattering data. It is understood that the imaginary amplitudes for small $|t|$ have
exponential forms $\exp(B_I(s)\, t/2)$ (with different slopes for pp and p\=p).
Derivation of the original DR forms with respect to $t$ leads to new relations.
With given energy dependence assumed  for the $B_I$ slopes, explicit PV calculations
can be performed, leading to predictions for the
derivatives of the real parts at $t=0$. If the energy dependence of the imaginary slopes is constructed
with a combination of power and logarithm  terms, similar to the PDG forms for
the total cross sections, the PV  are also  solved in terms of the Lerch's transcendent.
We thus arrive at analytical exact forms for the derivatives of the real amplitudes
 in the forward direction. This has enormous importance for the phenomenological
treatment of  pp and p\=p scattering.

The DRS can be used to investigate the structure of the real amplitudes in the forward range.
In accord with a theorem by A. Martin \cite{Martin}, at high energies it is observed
\cite{LHC7TeV, X1800, LHC8TeV} the existence of a zero that approaches the origin as the energy
increases.  This requires more than an exponential slope factor in the real part.
For example, assuming  the $t$ dependence of the real amplitude with
an exponential ($B_R\, t/2$) times a factor linear in the  $t$ variable, we may have
the expected zero. DRS predict a relation among the parameters, thus providing an
important theoretical control in the analysis of $d\sigma/dt$ data.

We stress that we are here limited to the short range strong interactions.
Coulomb interference is most important in the phenomenology of pp
and  p\=p scattering, and must be properly taken into account, but it is not included in
the concerns of the present work.

This paper  is organized as follows:

 In Section \ref{DRAS-section}  we review the
connections of imaginary and real parts of the amplitudes as given by general principles,
and write the forms of the dispersion relations for amplitudes and of the dispersion
relations for slopes in terms of principal value integrals of general forms.
In the  separate subsections
  \ref{DRA-section} and \ref{DRS-section},  the expressions for DR and DRS are
fully expanded, using the given inputs.

In Section \ref{Lerch} we present the proof of the analytical solution of the principal value
integrations in terms of elementary functions and Lerch's transcendents. General properties are
described, and explicit forms written for cases of practical occurrence. Cases of apparent
singularities are analyzed, and their cancellations  explained and explicitly
exhibited.

In Section \ref{DR_amplitudes_slopes}, with subsections for amplitudes and for derivatives,
we give explicit expressions
for the calculation of the real amplitude and of its derivative at $t=0$ in terms of the input
parameters of the total cross sections and of the imaginary slopes.

In Section \ref{Phenomenology} the connection of the mathematical results and the phenomenology
of pp and p\=p scattering is illustrated.

In Section \ref{conclusions} we list the  purposes and achievements of the present work.

Finally, in Appendix \ref{PV_Integrals_App} we proceed to a description of
properties of the principal value integrals from a more general point of view.

\section{ Dispersion Relations for Amplitudes and Slopes \label{DRAS-section}  }

The well known DR for pp and p\=p  elastic scattering are
written in terms
of even and odd dimensionless amplitudes,
\begin{eqnarray}
&&{\rm Re}\,F_+(E,t) = K +
\frac{2E^2}{\pi}\,{\bf P} \int_{m}^{+\infty}dE^\prime \, \frac{ {\rm Im}\, F_+(E^\prime,t)}{E^\prime(E^{\prime 2}-E^2)}\,,
\label{inteven}  \\
&&{\rm Re}\,F_-(E,t)=
\frac{2E}{\pi} \,{\bf P}\int_{m}^{+\infty}dE^\prime \, \frac{ {\rm Im}\, F_-(E^\prime,t)}{(E^{\prime 2}-E^2)} \, .
\label{intodd}
\end{eqnarray}
Here $E$ is the incident proton energy in lab system. The subtraction constant $K$
accounts for the convergence control in the one-subtracted DR.

In high energy processes the center of mass energy $\sqrt{s} $ is most commonly used. For pp and $\rm p \bar p$ scattering the
connection with the lab energy $E$ is
\begin{equation}
s=2mE+2m^2 ~,
\label{cm_energy}
\end{equation}
where $m$ is the proton/antiproton mass.
To work with the dispersion relations written above, the most useful quantity is the dimensionless
ratio
\begin{equation}
x=E/m
\label{x_variable}
\end{equation}
and then
\begin{equation}
\frac {s}{2m^2}= x+1 ~.
\label{s_x_relation}
\end{equation}
Approximate relations that are often used at high energies are obviously $s=2mE$ and $x=s/2m^2$.

The optical theorem informs the normalization of the amplitudes  by
\begin{eqnarray}
\sigma_{\rm pp}= \frac{{\rm Im} ~F_{\rm pp}(x,t=0)}{2m^2x}
\label{teorema_otico}
\end{eqnarray}
and similarly  for p\=p.

The even and odd combinations of amplitudes are related to the pp and p\=p systems through
\begin{equation}
F_{\rm pp}(x,t)=F_+(x,t)-F_-(x,t),   \qquad
F_{\rm {p \bar p} }(x,t)=F_+(x,t)+F_-(x,t) ~.
\label{F_pp}
\end{equation}

Assuming  for small $|t|$  exponential $t$ dependences for the imaginary parts of the amplitudes,
we write
\begin{eqnarray}
 && {\rm Im} ~F_{\rm pp}(x,t)= 2\,m^2\,x\,\sigma_{\rm pp}(x)\,\exp \left(B_I^{\rm pp}(x)\,t/2\right)\,,  \label{F_pp2} \\
 && {\rm Im} ~F_{\rm p\bar p}(x,t)= 2\,m^2\,x\,\sigma_{\rm p\bar p}(x)\,\exp \left(B_I^{\rm p\bar p}(x)\,t/2\right)\,,  \label{F_ppbar2}
\end{eqnarray}
with input functions $\sigma(x)$ and $B_I(x)$.
We obtain in this way for the even and odd inputs
\begin{eqnarray}
 &&{\rm Im} \,F_+(x,t)=  m^2\,x\Big[\sigma_{\rm p\bar p}(x)\,e^{B_I^{\rm p\bar p}(x)\,t/2}
 + \sigma_{\rm pp}(x)\,e^{B_I^{\rm pp}(x)\,t/2}\Big]\,,  \label{inputplus}  \\
 &&{\rm Im} \,F_-(x,t)=  m^2\,x\Big[\sigma_{\rm p\bar p}(x)\,e^{B_I^{\rm p\bar p}(x)\,t/2}
 - \sigma_{\rm pp}(x)\,e^{B_I^{\rm pp}(x)\,t/2}\Big]\,.
 \label{inputminus}
\end{eqnarray}
Substituting these expressions in Eqs.~(\ref{inteven}) and (\ref{intodd}), written in terms of the dimensionless variable $x$, we obtain
\begin{eqnarray}
&&{\rm Re}\,F_+(x,t) = K + \frac{2\,m^2\,x^2}{\pi}\,{\bf P}\!\int_{1}^{+\infty}\frac{1}{x^{\prime 2}-x^2}\,\Big[\sigma_{\rm p\bar p}(x^{\prime})\,\exp \left(B_I^{\rm p\bar p}(x^{\prime})\,t/2\right) \nonumber \\
 && \hspace{150pt} +\,\sigma_{\rm pp}(x^{\prime})\,\exp \left(B_I^{\rm pp}(x^{\prime})\,t/2\right)\Big]\,dx^\prime ,  \label{xinteven}  \\
&&{\rm Re}\,F_-(x,t) =  \frac{2\,m^2\,x}{\pi}\,{\bf P}\!\int_{1}^{+\infty}\frac{x^\prime}{x^{\prime 2}-x^2}\,\Big[\sigma_{\rm p\bar p}(x^{\prime})\,\exp \left(B_I^{\rm p\bar p}(x^{\prime})\,t/2\right)  \nonumber \\
 &&  \hspace{150pt} -\,\sigma_{\rm pp}(x^{\prime})\,\exp \left(B_I^{\rm pp}(x^{\prime})\,t/2\right)\Big]\,dx^\prime .
\label{xintodd}
\end{eqnarray}

The Particle Data Group  \cite{PDG} gives  parametrizations for the total cross sections
of the pp and p\=p interactions  in the well known forms
 \begin{eqnarray}
  \sigma^\mp(s)= P^{\prime}+ H^{\prime} \log^2\left(s/s_0\right)+R_1^{\prime} \left( s/s_0 \right)^{-\eta_1^{\prime}}
  \pm R_2^{\prime} \left( s/s_0 \right)^{-\eta_2^{\prime}},
  \label{PDG_sigma_s}
 \end{eqnarray}
with parameters $P^{\prime},~ H^{\prime}, ~ R_1^{\prime}, ~ R_2^{\prime}   $ in milibarns, $s_0$ in $\GeV^2$, and $\eta_1^{\prime} , ~\eta_2^{\prime}$
 dimensionless.  The upper and lower indices $-$, $+$   refer to p\=p
and pp scattering respectively. The parametrization is assumed to be adequate
for all energies $s \geq s_0$ .

 However, dispersion relations are defined with
respect to the lab system energy, and, for low energies, terms like $\log^2(E+m)$ and
$(E+m)^{-\eta}$ appear and spoil the simplicity of DR preventing to obtain  closed forms.
We then re-write (re-parametrize) the above values for the total cross sections in
terms of dimensionless variables
$x=E/m$, $x_0=E_0/m$,  with $x>1$, writing
\begin{eqnarray}
&& \sigma_{\rm pp}(x) = P+H\,\log^2 (x/x_0)+R_1\,(x/x_0)^{-\eta_1}-R_2\,(x/x_0)^{-\eta_2}\,,  \label{PDG sigma x1}  \\
&& \sigma_{\rm p\bar p}(x) = P+H\,\log^2 (x/x_0)+R_1\,(x/x_0)^{-\eta_1}+R_2\,(x/x_0)^{-\eta_2}\,,  \label{PDG sigma x2}
\end{eqnarray}
 and   obtain new parameters, with slight changes. Numerical values are given
in Section  \ref{Phenomenology}.
For mathematical simplicity, from now on we use in this paper the variable $x$ to represent the energy of
the collision, with use of the center of mass energy $\sqrt{s}$ in some places.

In terms of the  $x$ variable, the slopes $B_I^{\rm pp}(x)$  and $B_I^{\rm p\bar p}(x)$
are  written in the Regge-like forms
\begin{equation}
B_I^{\rm pp}(x)= b_0+b_1 \log(x) + b_2 \log^2(x)+b_3x^{-\eta_3} - b_4x^{-\eta_4} ~,
\label{BI_pp}
\end{equation}
\begin{equation}
B_I^{\rm p\bar{p}}(x)= b_0+b_1 \log(x) + b_2 \log^2(x)+b_3x^{-\eta_3} + b_4x^{-\eta_4} ~,
\label{BI_ppbar}
\end{equation}
with symmetry in the coefficients for pp and p\=p.
The suggested  numerical  values are given in Section \ref{Phenomenology}.

The even and odd inputs are given by Eqs.~(\ref{inputplus}, \ref{inputminus}).
Then the  DR for the PDG forms, Eqs.~(\ref{PDG sigma x1},  \ref{PDG sigma x2}), become
\begin{eqnarray}
&&{\rm Re}\,F_+(x,t) = K + \frac{2m^2 x^2 }{\pi}\,{\bf P}\int_{1}^{+\infty}
  \Bigg\{ \frac{P+ H \log^2({x^\prime}/{x_0})+R_1\, ({x^\prime}/{x_0})^{-\eta_1} }
          { x^{\prime 2}-x^2} \nonumber  \\
&& \hspace{180pt}    \times\,  \left(e^{B_I^{\rm p\bar{p}}(x^\prime)\,t/2}+e^{B_I^{\rm pp}(x^\prime)\,t/2}\right)  \nonumber \\
&&\hspace{50pt} + \,\frac{R_2\,({x^\prime}/{x_0})^{-\eta_2}}{ x^{\prime 2}-x^2}\left(e^{B_I^{\rm p\bar{p}}(x^\prime)\, t/2}-e^{B_I^{\rm pp}(x^\prime) \,t/2}\right)\Bigg\}\,dx^\prime
\label{xinteven2}
\end{eqnarray}
and
\begin{eqnarray}
&&  {\rm Re}\,F_-(x,t) =  \frac{2m^2 x}{\pi}\,{\bf P}\int_{1}^{+\infty} x^{\prime}\,\Bigg\{ \frac{P+ H \log^2({x^\prime}/{x_0})+R_1 \,({x^\prime}/{x_0})^{-\eta_1}}{ x^{\prime 2}-x^2} \nonumber  \\
&& \hspace{180pt} \times\,\left(e^{B_I^{\rm p\bar{p}}(x^\prime)\, t/2}-e^{B_I^{\rm pp}(x^\prime)\, t/2}\right)  \nonumber  \\
&&\hspace{50pt}+\, \frac{R_2 \,({x^\prime}/{x_0})^{-\eta_2}}{ x^{\prime 2}-x^2}\,\left(e^{B_I^{\rm p\bar{p}}(x^\prime)\, t/2}+e^{B_I^{\rm pp}(x^\prime) \,t/2}\right) \Bigg\} \,dx^\prime  \,.
  \label{xintodd2}
\end{eqnarray}

\subsection { Dispersion Relations for Amplitudes   \label{DRA-section}    }

Taking $t=0$ in Eqs. (\ref{xinteven}, \ref{xintodd})  we have
\begin{equation}
{\rm Re}\,F_+(x,0) = K + \frac{4m^2 x^2}{\pi}\,{\bf P}\int_{1}^{+\infty}
   \frac{P+ H \log^2({x^\prime}/{x_0})+R_1\, ({x^\prime}/{x_0})^{-\eta_1}   }
          { x^{\prime 2}-x^2}\, dx^\prime
\label{xinteven2}
\end{equation}
and
\begin{equation}
{\rm Re}\,F_-(x,0)=  \frac{4m^2 x}{\pi}\,
    {\bf P}\int_{1}^{+\infty}  \frac{x^{\prime}R_2\,({x^\prime}/{x_0})^{-\eta_2}}{ x^{\prime 2}-x^2}\, dx^\prime \,.
\label{xintodd2}
\end{equation}
In a glance at the integrands in these equations, we observe that both
 ${\rm Re}\,F_+(x,0)$ and ${\rm Re}\,F_-(x,0)$
result in linear combinations of PV integrals of the form
\begin{eqnarray}
I(n,\lambda,x)=
{\bf P}\int_{1}^{+\infty}\frac{x^{\prime\lambda} \, \log^n (x^\prime)}{ x^{\prime 2}-x^2}\,dx^\prime  ~,
\label{PV_int}
\end{eqnarray}
belonging to a family of integrals discussed in the Appendix. Exact values of these integrals, based on recent developments in
our study of the Lerch's transcendent \cite{EF-K-Sesma3}, can be given.

Collecting terms, we have for the even and odd parts
\begin{eqnarray}
&&{\rm Re}\,F_+(x,0) = K + \frac{4\,m^2\,x^2}{\pi} \,\Big[  I(0,0,x)\left(P+H \log^2 x_0\right)
  \nonumber \\
&&\hspace{25pt}+\,I(1,0,x) \left(-2 H \log x_0\right)+I(2,0,x)\, H
  + I(0,-\eta_1,x)\, R_1 \,x_0^{\eta_1} \Big]\quad
\label{even_even}
\end{eqnarray}
and
 \begin{eqnarray}
 {\rm Re}\,F_-(x,0) =   \frac{4\,m^2\,x}{\pi}\, I(0,1-\eta_2,x)  \, R_2 \,x_0^{\eta_2}  ~.
\label{odd_odd}
 \end{eqnarray}
Equations (\ref{even_even}) and (\ref{odd_odd}) are known DR relating imaginary and real parts of the complex
amplitude for   pp and   p\=p elastic scattering.

\subsection { Dispersion Relations for  Slopes   \label{DRS-section} }

For  small $|t|$ we extend the imaginary amplitude of the PDG representation introducing
factors $\exp[B_I^{\rm pp}(x)\, t/2]$  and  $\exp[B_I^{\rm p \bar p}(x)\, t/2]$, for  all terms in the
input form, as written above in Eqs.~(\ref{F_pp2}, \ref{F_ppbar2}).
The parametrizations of the imaginary slopes as functions of the energy allow us
to obtain, from the dispersion relations, information on the derivatives of the real parts at $|t|=0$.
This has essential importance for the construction of the forward  real amplitude, with
determination of the forward scattering parameters.

Taking derivatives of Eqs.~(\ref{xinteven}) and (\ref{xintodd})  with respect to $t$, we obtain
 \begin{eqnarray}
&& \frac{\partial{\rm Re}\,F_+(x,t)}{\partial t} = \frac{m^2\,x^2}{\pi}\,{\bf P}\!\int_{1}^{+\infty}
\frac{1}{x^{\prime 2}-x^2}\Big[\sigma_{\rm p\bar p}(x^{\prime})\,B_I^{\rm p\bar p}(x^{\prime})\,\exp \left(B_I^{\rm p\bar p}(x^{\prime})\,t/2\right) \nonumber \\
 && \hspace{100pt} +\,\sigma_{\rm pp}(x^{\prime})\,B_I^{\rm pp}(x^{\prime})\,\exp \left(B_I^{\rm pp}(x^{\prime})\,t/2\right)\Big]\,dx^\prime\,,  \label{A even}  \\
&& \frac{\partial {\rm Re}\,F_-(x,t)}{\partial t} = \frac{m^2\,x}{\pi}\,{\bf P}\!\int_{1}^{+\infty}
\frac{x^\prime}{x^{\prime 2}-x^2}\Big[\sigma_{\rm p\bar p}(x^{\prime})\,B_I^{\rm p\bar p}(x^{\prime})\,\exp \left(B_I^{\rm p\bar p}(x^{\prime})\,t/2\right)  \nonumber \\
 &&  \hspace{100pt} -\,\sigma_{\rm pp}(x^{\prime})\,B_I^{\rm pp}(x^{\prime})\,\exp \left(B_I^{\rm pp}(x^{\prime})\,t/2\right)\Big]\,dx^\prime\,.
\label{A odd}
\end{eqnarray}
which, with the parametrizations Eqs.~(\ref{PDG sigma x1}) to (\ref{BI_ppbar}), give for the dispersion relations
for the derivatives \cite{EF2007}  of the real amplitudes at the origin
 \begin{eqnarray}
\frac{\partial{\rm Re}\,F_+(x,t)}{\partial t}\Bigg|_{t=0}  &=&  \frac{2\,m^2\,x^2}{\pi}\, {\bf P}\int_{1}^{+\infty}\Bigg\{\Bigg(
   \frac{P+ H \log^2({x^\prime}/{x_0})+R_1\,({x^\prime}/{x_0})^{-\eta_1}}
          { x^{\prime 2}-x^2} \nonumber \\
&&\hspace{-100pt} \times\Big[b_0+b_1 \log(x^\prime)+b_2 \log^2(x^\prime)+b_3\,x'^{-\eta_3}\Big]\Bigg) + \frac{R_2\,(x'/x_0)^{-\eta_2}}{x^{\prime 2}-x^2}\,b_4\,x'^{-\eta_4}\Bigg\}  dx^\prime
\label{drdtplus}
\end{eqnarray}
and
\begin{eqnarray}
\frac{\partial {\rm Re}\,F_-(x,t)}{\partial t}\Bigg|_{t=0}  &=&  \frac{2\,m^2\,x}{\pi}\,{\bf P}\int_{1}^{+\infty}  \, x^{\prime} \, \Bigg\{
\frac{  P+ H \log^2({x^\prime}/{x_0})+R_1\,({x^\prime}/{x_0})^{-\eta_1} }{ x^{\prime 2}-x^2} \nonumber \\
&&\hspace{-100pt}\times\,b_4\,x'^{-\eta_4}+ \frac{ R_2\,({x^\prime}/{x_0})^{-\eta_2}   }
  { x^{\prime 2}-x^2} \,  \Big[b_0+b_1 \log(x^\prime)+b_2 \log^2(x^\prime)+b_3~x'^{-\eta_3}\Big]  \Bigg \} dx^\prime  \,.
\label{drdtminus}
\end{eqnarray}
The right-hand sides of these equations are also linear combinations of PV integrals of the form $I(n,\lambda,x)$ defined in Eq.~(\ref{PV_int}).

Collecting terms in these equations, we obtain for the even part
\begin{eqnarray}
&& \frac{\partial{\rm Re}\,F_+(x,t)}{\partial t}\Bigg|_{t=0} =
 \frac{2\,m^2\,x^2}{\pi}\, \Bigg\{ I(0,0,x)~  \Big(P+H \log^2{x_0}\Big)~  b_0    \nonumber \\
&& \hspace{75pt} +\,I(1,0,x) \Big[ \Big(-2 H \log{x_0}\Big)~ b_0  + \Big(P+H \log^2{x_0}\Big)~ b_1 \Big]  \nonumber \\
&& \hspace{75pt} +\,I(2,0,x)\Big[  H ~ b_0
 -2 H \log{x_0} ~  b_1 + \Big(P+H \log^2 {x_0} \Big)~ b_2  \Big]   \nonumber  \\
&& \hspace{75pt}+\,I(3,0,x)  ~ \Big[ -2 H \log{x_0} ~   b_2    +  H ~ b_1 \Big] +I(4,0,x) ~     H ~  b_2 \nonumber \\
&& \hspace{40pt} + \,  R_1 ~ x_0^{\eta_1}  \Big(I(0,-\eta_1,x)~b_0+I(1,-\eta_1,x)~b_1+I(2,-\eta_1,x)~ b_2 \nonumber \\
&& \hspace{40pt}+\,I(0,-\eta_1-\eta_3,x)~b_3\Big)+ R_2~x_0^{\eta_2}~I(0,-\eta_2-\eta_4,x)~ b_4       \nonumber \\
&&\hspace{40pt}+\, \Big[(P+H\log^2 {x_0})\,I(0,-\eta_3,x)-2H\log{x_0}\,I(1,-\eta_3,x)  \nonumber  \\
&& \hspace{180pt}+\,H~I(2,-\eta_3,x)\Big] ~b_3 \Bigg\} ,
 \label{drdtplus_expanded}
\end{eqnarray}
and for the odd part
\begin{eqnarray}
&& \frac{\partial{\rm Re}\,F_-(x,t)}{\partial t}\Bigg|_{t=0} =
 \frac{2\,m^2\,x}{\pi}\, \Bigg\{ R_2 ~ x_0^{\eta_2}\Big(I(0,1-\eta_2,x)~ b_0 + I(1,1-\eta_2,x)~ b_1
    \nonumber \\
   &&\hspace{100pt} +\,I(2,1-\eta_2,x) ~ b_2+ I(0,1-\eta_2-\eta_3,x)~b_3 \Big)  \nonumber \\
&&\hspace{50pt} +\,\Big( (P+H\log^2{x_0})\,I(0,1-\eta_4,x)
-2H\,\log{x_0}~I(1,1-\eta_4,x)  \nonumber \\
&& \hspace{50pt}+\,H~I(2,1-\eta_4,x)
+R_1~x_0^{\eta_1}~I(0,1-\eta_1-\eta_4,x)\Big)  ~ b_4~\Bigg\}~.
\label{drdtminus_expanded}
\end{eqnarray}

These equations, here called Dispersion Relations for Slopes, control quantities observed in
forward scattering and should be used as basic information for
phenomenological and  theoretical description of
forward pp and $\rm{ p \bar p}$ scattering. In their introduction \cite{EF2007}, they were shown to be important for the
analysis of forward scattering, determining structure and parameters of the real amplitude.

The next section shows our procedure for calculating the PV integrals of the type $I(n,\lambda,x)$ entering
 Eqs.~(\ref{even_even}, \ref{odd_odd}, \ref{drdtplus_expanded}, \ref{drdtminus_expanded}).

\section { \label{Lerch} Principal Value Integrals and  Lerch's transcendent}

The PV integrals have always had  a protagonist role in the  applications of the
principles of DR to pp and p\=p scattering, requiring   much  work and concern.
Integrands with singularities were not easy to deal with, specially before powerful numerical
computation methods  became   accessible. Dynamical details, such as resonances and thresholds,
were  mixed up  in the efforts to obtain  the real part of the amplitude. It was not
easy to  separate the physical details from the mathematical difficulties.

Nowadays we can show that in high energy pp and p\=p scattering, analytically very  simple inputs,
treated with direct exact mathematics, can account for all observation, in high precision.
The simplified mathematics helps to show that the phenomenology of the physical structure
is also  very simple. Existing physical complications are not visible in the analysis
of the data, within the existing experimental precision.

First, the input form of the imaginary amplitude, containing only powers and
logarithms, was
treated by the DDR, avoiding the need  of numerical integrations through
singularities. These relations allowed the identification of some global properties of
the observable quantities.
The limitations of the first DDR forms, restricted to high energies, were solved
with exact expressions \cite{Menon}, using double series. The difficulties with
 proofs on convergence of double infinite series required  numerical proofs
comparing principal value integrations with  series summations.
Further progress came with  the reduction of the representations to single
infinite series, of proved convergence \cite{EF-Sesma1}.  The convergence of
the series is both mathematically sure, and    fast and comfortable  in practice.
Later, another construction of the representation of the PV integrals was introduced
\cite{EF-Sesma2}, identifying the  treatment and cure of singularities that occur
for certain values of parameters, showing clearly their  cancellation.
Very recently, further formal progress came with the study of properties of
the Lerch's transcendent. The proof given \cite{EF-K-Sesma3}
for a new representation  of the so called Lerch's transcendent
allows to express the PV integrals of hadronic phenomenology  in
compact and closed form, in terms of  these well studied functions of the
mathematical  literature. The  theorem is reproduced below, together with
the proof that it contains the method to write the solutions of the PV
that appear in DR and DRS.

The Lerch's transcendent $\Phi(z,s,a)$ (see Sec.~25.14 of Ref.~11), also called  Hurwitz-Lerch zeta function,
 is defined by its series representation
 \begin{equation}
\Phi(z,s,a) = \sum_{m=0}^\infty\frac{z^m}{(a+m)^s}\,,
\label{i1}
\end{equation}
with
\begin{equation}
a\neq 0, -1, -2, \ldots;\qquad \qquad |z|<1;  \qquad  |z|=1, \quad  {\rm Re}\,(s) >1\,.
\label{i2}
\end{equation}
The restriction on the values of $a$ guarantees that all terms of the series in the right-hand side are finite. Obviously,
the series is convergent if $|z|<1$, independently of the value of $s$, or if $|z|=1$ and ${\rm Re}\,s>1$.
For other values of its arguments, $\Phi(z,s,a)$ is defined by analytic continuation, that is achieved by
means of integral representations.
Characteristics of the $\Phi$ function are  the identities
\begin{eqnarray}
\Phi(z,s,a+1) & = & \frac{1}{z}\left(\Phi(z,s,a)-\frac{1}{a^s}\right),   \label{bat2}  \\
\Phi(z,s-1,a) & = & \left(a+z\frac{\partial}{\partial z}\right)\Phi(z,s,a)\,,      \label{bat3}  \\
\Phi(z,s+1,a) & = & -\frac{1}{s}\,\frac{\partial}{\partial a}\Phi(z,s,a)\,,     \label{bat4}
\end{eqnarray}
stemming from the series representation  in Eq.~(\ref{i1}).

A  new representation  of   $\Phi(z,n,a)$  that establishes its connection with the
PV integrals of the theory of DR   was  recently  proved \cite{EF-K-Sesma3} with the following

\noindent{\bf Theorem.  }
{\it
Let  $z$ be  a complex number belonging to the open disc of radius 1, excluded its center at the
origin, and cut along the negative real semiaxis, that is,
\begin{equation}
z\in \mathcal{C}, \qquad 0<|z|<1, \qquad -\pi<\arg(z)<\pi.
\label{ii1}
\end{equation}
Let us denote
\begin{equation}
\varphi=\arg(-\log z)  \label{ii2}  \,.
\end{equation}
Then, for positive integer values of $n=1, 2, \ldots$ and for complex $a$ such that ${\rm Re}\,[(a-1)e^{ i\varphi}]<0$,
the Lerch's transcendent admits the representation
 \begin{equation}
\Phi(z,n,a) = \frac{(-1)^{n-1}}{(n-1)!}
\left\{
{\bf P} \int_0^{\infty e^{i\varphi}}\frac{t^{n-1}\,e^{at}}{z\,e^t-1}\,dt
 + \pi\,\frac{\partial^{n-1}}{\partial a^{n-1}}\left(z^{-a}\,\cot(\pi a)\right)
 \right\},   \label{ii3}
\end{equation}
where the symbol ${\bf P}$ stands for the Cauchy principal value of the path
integral along the ray $\arg(t)=\varphi$.  }

With specifications for particular cases, the theorem is applied to obtain the expressions
for the general  PV integrals in Eq.~(\ref{PV_int}) that we need.
 Assuming that $z$ and $a$ are real,   then
it is $\varphi =0$. Changing  the integration variable with  $t=2\,\log (x^{\prime})$,
we may write
\begin{eqnarray}
&&\Phi(z,n+1,a)= \frac{(-1)^{n}}{(n)!} \bigg[ {\bf P}\int_1^{+\infty} x^{\prime (2 a-1)}
\frac{2^{(n+1)} \log^n(x^\prime)}{z \,x^{\prime 2}-1} \, dx^\prime \nonumber  \\
&&\hspace{130pt} +\, \pi\,\frac{\partial^{n}}{\partial a^{n}}\left(z^{-a}\,\cot(\pi a)\right) \bigg]  ~ .
\label{ii3_2}
\end{eqnarray}
Putting $z=1/x^2$ and $a=(1+\lambda)/2$ we finally obtain the  general form for the PV integrals,
defined by Eq.~(\ref{PV_int}),
\begin{eqnarray}
&&I(n,\lambda ,x) =\frac{\pi}{2x}\frac{\partial ^n}{\partial \lambda^n}\left[x^{\lambda} \tan \left(\frac{\pi}{2}\lambda\right)\right] \nonumber  \\
&&\hspace{70pt}+\,\frac{(-1)^n}{x^2} 2^{-(n+1)}\, n!  ~  \Phi\left(\frac{1}{x^2},n+1,\frac{1+\lambda}{2}\right) ~ .
\label{PV_integrals_app}
\end{eqnarray}
Thus, using the  recent developments \cite{EF-K-Sesma3}, we have obtained general exact  forms
of the PV integrals $I(n,\lambda,x)$ with non-negative integer $n$, complex $\lambda$ such that ${\rm Re}\,\lambda<1$,
and real $x>1$. Under these conditions, the function $\Phi$  provides elegant exact representations for the PV integrals of the DR.
Equation (\ref{i1}) allows us to write for the function $\Phi$ in the right-hand side of  Eq.~(\ref{PV_integrals_app}) the expansion
\begin{equation}
\Phi\left(\frac{1}{x^2}, n+1,\frac{1 + \lambda}{2}\right)
  = 2^{n+1}\sum_{j=0}^{\infty} \frac{x^{-2j} }{(2j+1 + \lambda)^{n+1}}\,.
\label{Lerch_app}
 \end{equation}
For its derivative with respect to $\lambda$, use can be made of Eq.~(\ref{bat4}) to write
 \begin{equation}
 \frac{\partial}{\partial \lambda}\Phi \left(\frac{1}{x^2},n+1,\frac{1+\lambda}{2}\right) = -\,\frac{n+1}{2} ~ \Phi\left(\frac{1}{x^2},n+2,
\frac{1+\lambda}{2} \right) ~.
\label{recurr_app}
\end{equation}
In the applications to pp and ${\rm p \bar p }$ scattering, the sums converge rapidly
for energies  above $\sqrt{s} \approx 10 \GeV$, and are easily included in practical computations,
requiring  only one or a few terms of the series.

We remark that $ I(n=0,\lambda=0 ,x)$ can be written in terms of elementary functions,
\begin{equation}
I(0,0,x) =
   {\bf P}\!\int_{1}^{+\infty}\!\frac{1  }
          { x^{\prime 2}-x^2}~ dx^\prime  = \frac{1}{2\, x^2} ~  \Phi\left(\frac{1}{x^2},1,\frac{1}{2}\right)
= \frac{1}{2\,x }\,\log \frac{x+1}{x-1}  ~ .
\label{I00}
 \end{equation}

We may write  a simple combination that eliminates the denominator
in the PV integrals to obtain
\begin{eqnarray}
I(n,\lambda,x)- x^2 ~ I(n,\lambda-2,x)&=& \int_{1}^{+\infty}
  x^{\prime (\lambda-2)} \log^n(x^\prime)  dx^\prime   \nonumber  \\
  &=& (1-\lambda)^{-1-n}\, \Gamma(1+n) ~ ,
\label{comb}
\end{eqnarray}
that does not depend on  $x$.
On the other hand, using the property of periodicity of  the
tangent function  in Eq.~(\ref{PV_integrals_app}),
the combination eliminates the terms with derivatives and can also be written
\begin{eqnarray}
I(n,\lambda,x)- x^2 ~ I(n,\lambda-2,x) &=&   \frac{(-1)^n}{x^2} 2^{-(n+1)}\Gamma(1+n) \nonumber  \\
&& \hspace{-60pt}\times\,\left[
\Phi\left(\frac{1}{x^2},n+1,\frac{1+\lambda}{2}\right)-x^2 \Phi\left(\frac{1}{x^2},n+1,\frac{1+\lambda-2}{2}\right)\right]  \nonumber  \\
 &=& (1-\lambda)^{-1-n} \,\Gamma(1+n)  ~ ,
\label{comb_Lerch}
\end{eqnarray}
where in the last step use has been made of the property of the Lerch's transcendent in
Eq.~(\ref{bat2}).
This confirms the  expression  for the general formula  in Eq.~(\ref{PV_integrals_app})  for the
 PV integration.
The combination  free of derivatives can be used for
non-integer $n$, and can also be useful for computational purposes, since full
calculations of   integrals  need to be made only for $\lambda\in (-1,\, 1)$.

Specific values for the PV integrals used in this work
(values of $n=0,1,2,3,4$) are given below .
\begin{eqnarray}
I(0,\lambda ,x)&=&\frac{\pi}{2}~ x^{\lambda-1} \tan \Big(\frac{\pi \lambda}{2}\Big) +
\frac{1}{2 x^2}~ \Phi\Big(\frac{1}{x^2},1,\frac{1+\lambda}{2}\Big)  ~ ,
\label{In0}  \\
I(1,\lambda ,x) &=& \frac{\pi}{2}~ x^{\lambda-1} \left( \log(x)  \tan \Big(\frac{\pi \lambda}{2}\Big) +
\frac{\pi}{2} \sec^2\Big(\frac{\pi \lambda}{2}\Big) \right)\nonumber \\
&& \hspace{120pt} - \frac{1}{4 x^2} ~ \Phi\Big(\frac{1}{x^2},2,\frac{1+\lambda}{2}\Big)
  ~ ,   \label{In1}  \\
I(2,\lambda ,x) &=& \frac{\pi}{2}~ x^{\lambda-1} \Bigg[ \log^2(x)  \tan \Big(\frac{\pi \lambda}{2}\Big)
+ \pi \sec^2\Big(\frac{\pi \lambda}{2}\Big) \Bigg( \log(x)
 \nonumber \\
&& \hspace{50pt}+\, \frac{\pi}{2}\tan \Big(\frac{\pi \lambda}{2}\Big)\Bigg) \Bigg]+  \frac{1}{4x^2} ~ \Phi\Big(\frac{1}{x^2},3,\frac{1+\lambda}{2}\Big)
  ~ . \label{In2}  \\
I(3,\lambda ,x) &=& \frac{\pi}{2}~ x^{\lambda-1} \left\{ \log^3(x)  \tan \Big(\frac{\pi \lambda}{2}\Big)+ \frac{\pi}{2}
\sec^2\Big(\frac{\pi \lambda}{2}\Big) \Bigg[ 3 \log^2(x)\right.\nonumber  \\
 &&\hspace{-90pt}\left. + \,3 \pi \log(x) \tan \Big(\frac{\pi \lambda}{2}\Big) + \frac{\pi^2}{2} \left(1+ 3 \tan^2\Big(\frac{\pi \lambda}{2}\Big)\right)   \Bigg] \right\}
 - \frac{3}{8x^2} \,\Phi\Big(\frac{1}{x^2},4,\frac{1+\lambda}{2}\Big)\, ,
 \label{In3}  \\
I(4,\lambda ,x) &=& \frac{\pi}{2}~ x^{\lambda-1} \left\{ \log^4(x) \tan \Big(\frac{\pi \lambda}{2}\Big)
  + \pi \sec^2\Big(\frac{\pi \lambda}{2}\Big) \bigg[ 2  \log^3(x)  \right. \nonumber  \\
 && \hspace{5pt} + \, 3 \pi \log^2(x) \tan \Big(\frac{\pi \lambda}{2}\Big)  + \pi^2  \log(x) \left(1 + 3  \tan^2 \Big(\frac{\pi \lambda}{2}\Big)\right)   \nonumber  \\
&& \hspace{-40pt}\left.+ \frac{\pi^3}{2}\tan \Big(\frac{\pi \lambda}{2}\Big)\left(2+3 \tan^2\Big(\frac{\pi \lambda}{2}\Big)\right) \bigg] \right\}  +\frac{3}{4x^2} ~  \Phi\left(\frac{1}{x^2},5,\frac{1+\lambda}{2}\right) .  \label{In4}
\end{eqnarray}

The use of Eq.~(\ref{PV_integrals_app}) is straightforward, except that care must be
taken for odd negative integer values of $\lambda=-(2N+1)$,
with $N$ zero or positive integer, when singularities
occur in both  trigonometric  and $\Phi$ function parts of the expression,
 with cancellation in a limit procedure that has been explained before \cite{EF-Sesma2}.
Examples of the calculation
of the limits are given below.
\begin{eqnarray}
 I(0,-(2N+1),x) &=& x^{-2N-2}\Big[-\log(x)  -\frac{1}{2}\sum_{k=1}^N\frac{x^{2k}}{k}+\frac{1}{2}\,{\rm Li}_1(x^{-2})\Big],
\label{limits_1}  \\
 I(1,-(2N+1),x)  &=&  x^{-2N-2}\Big[-\frac{1}{2}\log^2(x)   +\frac{\pi^2}{12}-\,\frac{1}{4}\sum_{k=1}^N\frac{x^{2k}}{k^2}  \nonumber  \\
 &&\hspace{120pt}-\,\frac{1}{4}\,{\rm Li}_2(x^{-2})\Big],
\label{limits_2}  \\
I(2,-(2N+1),x)  &=&  x^{-2N-2}\Big[-\frac{1}{3}\log^3(x) +\frac{\pi^2}{6}\log(x) -\,\frac{1}{4}\sum_{k=1}^N\frac{x^{2k}}{k^3}\nonumber \\
&& \hspace{120pt}+\,\frac{1}{4}\,{\rm Li}_3(x^{-2})\Big],
\label{limits_3}   \\
 I(3,-(2N+1),x) &=&  x^{-2N-2}\Big[-\frac{1}{4}\log^4(x)  +\frac{\pi^2}{4}\log^2(x) +\frac{\pi^4}{120}  \nonumber  \\
&&  \hspace{80pt}-\,\frac{3}{8}\sum_{k=1}^N\frac{x^{2k}}{k^4} - \frac{3}{8}\,{\rm Li}_4(x^{-2})\Big],
\label{limits_4}  \\
 I(4,-(2N+1),x) &=&  x^{-2N-2}\Big[-\frac{1}{5}\log^5(x)+\frac{\pi^2}{3}\log^3(x)+\frac{\pi^4}{30}\log(x)  \nonumber   \\
&& \hspace{80pt}-\,\frac{3}{4}\sum_{k=1}^N\frac{x^{2k}}{k^5} +\frac{3}{4}\,{\rm Li}_5(x^{-2})\Big],
\label{limits}
\end{eqnarray}
where ${\rm Li}_m$ represents the polylogarithm of order $m$.
We recall  that the polylogarithm functions that  appear above are  defined by the simple series
 \begin{equation}
 {\rm Li}_m(x^{-2})=\sum_{k=1}^\infty \frac{x^{-2k}}{k^m} ~ .
\label{polylog}
\end{equation}

\section { \label{DR_amplitudes_slopes} Real Parts of Amplitudes and Derivatives from DR and DRS}

Based on the inputs  of total cross sections  $\sigma(x)$ for pp and  p\=p,  DR determine  the real amplitudes at $t=0$ as
functions of the energy. The subtraction constant $K$ is required
(obtained from data), with a unique energy independent value. With basis on the input of the imaginary slope $B_I(x)$
for pp and p\=p, DRS determine the derivatives of the real amplitudes at $t=0$ as functions of the energy.

With terms of general form  $x^\lambda \log^n(x)$  in the inputs, exact solutions are written for all  PV integrals
that appear in DR and DRS. Thus exact forms, valid for all energies,  are written for the real amplitudes and for their derivatives
in the forward direction. Since it is known that the real part has important structure in the forward range, these
results give essential contributions to the analysis of the dynamics governing elastic processes.
Below we provide practical expressions for the results of DR and DRS, keeping the dominant terms of the Lerch's transcendents.
In Section \ref{Phenomenology} we illustrate  the use of these  results in  the analysis of scattering data.

 \subsection{Real Amplitudes at $t=0$  \label{amplitudes_DR} }

For practical use, taking the   low energy corrections to first order, we write
 below the  expressions for the $\rho \sigma$ products obtained with the PDG (imaginary
 amplitude)  input.   We have for the even part
 \begin{eqnarray}
 \frac{1}{2}\Big[(\sigma\rho) ({\rm p \bar p})+(\sigma\rho) ({\rm pp})\Big] &=&
  \frac{1}{2 m^2 x}\,{\rm Re}~ F_+(x,0)  \nonumber \\
   &=&   T_1(x)+T_2(x)+T_3(x) ,   \label{rhosigma_terms}
\end{eqnarray}
with
\begin{eqnarray}
&& T_1(x)= H \pi \log\Big(\frac{x}{x_0}\Big) ~ ,
\label{T1-eq}  \\
&& T_2(x) = \frac{K}{2 m^2 \,x}+\frac{2}{ \pi\,x } \Big( P+H \left[  \log ^2\left(x_0\right)
+2 \log\left(x_0 \right) +2   \right] \Big) \, ,
 \label{T2-eq}  \\
&& T_3(x)=   R_1 \,x_0^{\eta_1} \Big[  -\,x^{-\eta_1}\, \tan \Big(\frac{\pi\, \eta_1}{2}\Big)
 +    \frac{1}{x} ~ \frac{2/\pi}{1-\eta_1} \Big] ~,
\label{T3-eq}
\end{eqnarray}
and for the odd part
\begin{eqnarray}
 \frac{1}{2}\left[(\sigma\rho) ({\rm p\bar p})-(\sigma\rho) ({\rm p  p}) \right]
&=&\frac{1}{2 m^2x}{\rm Re}\,F_-(x,0)  \nonumber  \\
 &=&   R_2\,x_0^{\eta_2} \Big[x^{-\eta_2}\, \cot\Big(\frac{\pi\, \eta_2}{2}\Big)
 +     \frac{1}{x^2} ~  \frac{2/\pi}{2-\eta_2}   \Big].
\label{T_odd-eq}
\end{eqnarray}
Additional  terms are of order ${\cal O}(x^{-4})$.

\subsection { \label{slopes_DRS} Derivatives of Real Amplitudes at $t=0$ }

We can learn more about the  $|t|$  dependence of the amplitudes through the
investigation of the DR for slopes, DRS.
For practical  purposes we give below explicit expressions for the DRS    including
only the first  term of the expansion of the transcendents. These forms are sufficient for high energies
(say $\sqrt{s} > 200 \GeV $). We have
 \begin{eqnarray}
\frac{1}{2m^2 x}\frac{\partial{\rm Re}\,F_+(x,t)}{\partial t}\Big|_{t=0} &=&  \nonumber  \\
&&\hspace{-100pt} \frac{ 1}{\pi} ~ \Big[(P+H\log^2 x_0)~G_1(x)+H~G_2(x)+R_1~G_3(x)+R_2~G_4(x)\Big],
 \label{drdtplus_expanded_1}
\end{eqnarray}
where
\begin{eqnarray}
G_1(x)&\equiv& \frac{b_0-b_1+2b_2}{x}+\frac{b_1 \pi^2}{4}+\frac{b_2 \pi^2}{2}\log x  \nonumber  \\ &&\hspace{75pt}+\,b_3\bigg(-\frac{\pi}{2}x^{-\eta_3}\tan\left(\frac{\pi\eta_3}{2}\right)+\frac{1}{x}\,\frac{1}{1-\eta_3}\bigg)~,
\label{G_1}  \\
G_2(x)&\equiv& \bigg[\frac{\pi^2}{4}\Big(3\log^2 x+\frac{\pi^2}{2}\Big)-\frac{6}{x}\bigg](b_1-2b_2\log x_0)-2b_0\log x_0\Big(\frac{\pi^2}{4}-\frac{1}{x}\Big) \nonumber \\
&&\hspace{-20pt}+\,(b_0-2b_1\log x_0)\Big(\frac{\pi^2}{2}\log x+\frac{2}{x}\Big)
+b_2\bigg[\pi^2\log x\Big(\log^2x+\frac{\pi^2}{2}\Big)+\frac{24}{x}\bigg]
\nonumber \\
&&\hspace{-20pt}-\,\pi\, b_3 \,x^{-\eta_3}\bigg[\log x~\tan\Big(\frac{\pi\eta_3}{2}\Big)\Big(-\log x_0+\frac{1}{2}\log x\Big)
-\frac{\pi}{2}\sec^2 \Big(\frac{\pi\eta_3}{2}\Big)\nonumber \\
&&\hspace{-20pt}\times\,\bigg(\log \Big(\frac{x}{x_0}\Big)-\,\frac{\pi}{2}\tan\left(\frac{\pi\eta_3}{2}\right)\bigg)\bigg]
+\frac{2b_3}{x(1-\eta_3)^2}\Big(\log x_0+\frac{1}{1-\eta_3}\Big),
\label{G_2}  \\
G_3(x)&\equiv& x_0^{\eta_1}\bigg\{
b_0\bigg[\frac{\pi}{2}x^{-\eta_1}\tan\left(-\frac{\pi\eta_1}{2}\right)+\frac{1}{x}\,\frac{1}{1-\eta_1}\bigg] \nonumber   \\
&&+\,b_1\,\frac{\pi}{2}\,x^{-\eta_1}\bigg[\frac{\pi}{2}\sec^2\left(\frac{\pi\eta_1}{2}\right)
-\tan\left(\frac{\pi\eta_1}{2}\right)\log x \bigg] \nonumber \\
&&\hspace{-40pt}-\,b_2\,\frac{\pi^2}{2}\,x^{-\eta_1}\bigg[\sec^2\left(\frac{\pi\eta_1}{2}\right)\bigg(\frac{\pi}{2}\tan\left(\frac{\pi\eta_1}{2}\right) -\log x\bigg) +\frac{1}{\pi}\tan\left(\frac{\pi\eta_1}{2}\right)\log^2 x \bigg]  \nonumber \\
&&+\,b_3\bigg[-\frac{\pi}{2}x^{-\eta_1-\eta_3}\tan\Big(\frac{\pi(\eta_1+\eta_3)}{2}\Big)
+\frac{1}{x}\,\frac{1}{1-\eta_1-\eta_3}\bigg]  \nonumber  \\
&&\hspace{80pt}+\,\frac{1}{x}\,\frac{1}{(1-\eta_1)^2}\bigg(-b_1+\frac{2 b_2}{(1-\eta_1)}\bigg)\bigg\}  ~,
\label{G_3}  \\
G_4(x)&\equiv& x_0^{\eta_2}\,b_4\,\left[-\frac{\pi}{2}x^{-\eta_2-\eta_4}\tan\Big(\frac{\pi(\eta_2+\eta_4)}{2}\Big) +\frac{1}{x}\,\frac{1}{1-\eta_2-\eta_4}\right]~.
\label{G_4}
\end{eqnarray}
For the odd combination we have
 \begin{eqnarray}
\frac{1}{2m^2 x}\frac{\partial{\rm Re}\,F_-(x,t)}{\partial t}\Big|_{t=0} &=&  \nonumber \\
&& \hspace{-100pt}\frac{1}{\pi}~ \Big[(P+H\log^2 x_0)~F_1(x)+H~F_2(x)
+R_1~F_3(x)+R_2~F_4(x)\Big],
\label{drdtminus_expanded_1}
\end{eqnarray}
where
\begin{eqnarray}
F_1(x)&\equiv& b_4\bigg(\frac{\pi}{2}x^{-\eta_4}\cot\left(\frac{\pi\eta_2}{2}\right)+\frac{1}{x^2}\,\frac{1}{2-\eta_4}\bigg)~,
\label{F_1}  \\
 F_2(x)&\equiv& b_4\bigg\{\frac{\pi}{2}\,x^{-\eta_4}\bigg[\pi\csc^2\Big(\frac{\pi\eta_4}{2}\Big)\bigg(\log \Big(\frac{x}{x_0}\Big)+
\frac{\pi}{2}\cot\Big(\frac{\pi\eta_4}{2}\Big)  \bigg)    \nonumber \\
&&\hspace{-50pt}+\,\cot\Big(\frac{\pi\eta_4}{2}\Big)\log x\,(-2\log x_0+\log x)  \bigg]+\frac{2}{x^2}\,\frac{1}{(2-\eta_4)^2}\Big(\log x_0+\frac{1}{2-\eta_4}\Big)  \bigg\},
\label{F_2}  \\
F_3(x)&\equiv& b_4\,x_0^{\eta_1}\bigg[\frac{\pi}{2}x^{-\eta_1-\eta_4}\cot\Big(\frac{\pi}{2}(\eta_1+\eta_4)\Big)
+\frac{1}{x^2}\,\frac{1}{2-\eta_1-\eta_3} \bigg]~,
\label{F_3}  \\
F_4(x)&\equiv& x_0^{\eta_2}\bigg\{\frac{\pi}{2}x^{-\eta_2}\bigg[ \cot\left(\frac{\pi\eta_2}{2}\right)\left(b_0+b_1\log x+b_2\log^2 x\right) \nonumber \\
&&\hspace{-50pt}+\,\frac{\pi}{2}\csc^2\left(\frac{\pi\eta_2}{2}\right)\bigg(b_1+\pi\cot\left(\frac{\pi\eta_2}{2}\right)b_2+2\log x~b_2 \bigg)
+x^{-\eta_3}~\cot\Big(\frac{\pi}{2}(\eta_2+\eta_3)\Big)b_3 \bigg]    \nonumber \\
&&\hspace{30pt}+\,\frac{1}{x^2}\frac{1}{2-\eta_3}\bigg[b_0-\frac{b_1}{2-\eta_2}+\frac{2 b_2}{(2-\eta_2)^2}+\frac{(2-\eta_2)b_3}{2-\eta_2-\eta_3}  \bigg]\bigg\}~,
\label{F_4}
\end{eqnarray}

\section {Relation with Phenomenology      \label{Phenomenology} }

We are concerned  with  the strong interaction part of pp and p\=p  scattering,
with no consideration for the Coulomb interference. We stress that the proper
treatment of the real part is  essential for a correct account of the Coulomb interference.
The determinations of the total cross section, through its connection with the optical point
of the imaginary amplitude, and of the $\rho$ parameter
of the ratio of real   to imaginary amplitudes at $t=0$, through analysis of  $d\sigma/dt$ data,
are affected by the structure (namely the $t$ dependence) of the real part and its
interference with the Coulomb amplitude.
DR and DRS determine values of the real amplitude and its derivative at the origin,
controlling parameters of the real and imaginary parts.

In this paper we prove that, in DR and DRS, terms of the general form  $ x^\lambda \log^n(x)$  with integer $n$
 in amplitudes and   slopes of  the   imaginary part   can receive  exact treatment,
through analytic  solution of the intervening PV integrals.
Luckily,  the well known and well accepted parametrization of the  pp and p\=p total cross sections made
by  the PDG/Compete Collaboration,  is a linear combination of such terms. With
parametrization of the imaginary slopes $B_I$ also made with such terms, the
DRS also come with the exactly calculable forms. Thus, we obtain the new
results presented in this work.  With knowledge of    exact explicit forms
for DR and DRS, we have powerful  support for the analysis of forward $d\sigma/dt $ data.

The parametrization of the Particle Data Group/Compete Collaboration  \cite{PDG}  for the total cross section
for  pp and p\=p interaction  in the   form  of Eq.~(\ref{PDG_sigma_s}) with functions of $s$
has parameter values
  $P^{\prime},~ H^{\prime}, ~ R_1^{\prime}, ~ R_2^{\prime}   $ in milibarns, $s_0$ in $\GeV^2$,
and $\eta_1^{\prime} , ~\eta_2^{\prime}$  dimensionless. The parametrization is considered  adequate
for all energies $s \geq s_0$ . Since  dispersion relations are written with
  the lab system energy,  we re-parametrize the   above form,
writing similar   Eqs.~(\ref{PDG sigma x1}, \ref{PDG sigma x2})  in terms of the
variable $x$, and find parameter values
  $P=34.37 $ mb,  $H=0.2704$ mb, $R_1=12.46$ mb, $R_2=7.30$ mb,
$\eta_1=0.4258$, $\eta_2=0.5458$ and $x_0= 8.94$.

Similarly, the experimental $B_I$ slopes are represented by the forms of
  Eqs.~(\ref{BI_pp}, \ref{BI_ppbar}),
with suggested  parameter values
 $b_0 = 13.03$ GeV$^{-2}$, $b_1=-0.3346$ GeV$^{-2}$, $b_2=0.04255$ GeV$^{-2}$, $b_3=-6.94$ GeV$^{-2}$,
$b_4=17.31$ GeV$^{-2}$, $\eta_3=0.5154$ and
$\eta_4=0.960$. The symmetries in the expressions for pp and p\=p simplify the algebra,
without   loss in the quality of the representations of the data.

\medskip
\noindent{\bf The Subtraction Constant and the Parameter $\rho$. }

\begin{figure*}[b]
 \includegraphics[width=8cm]{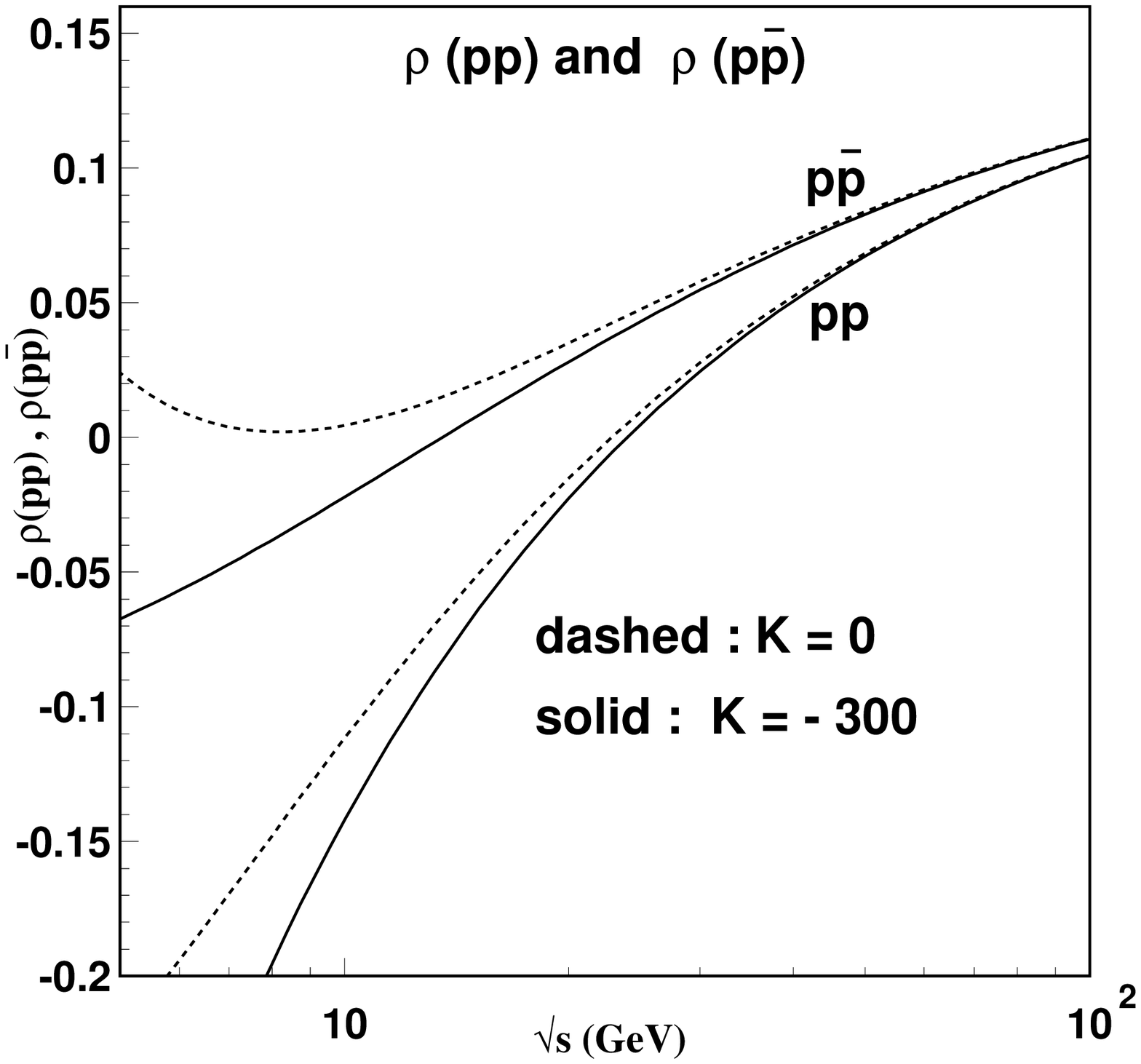}
  \includegraphics[width=8cm]{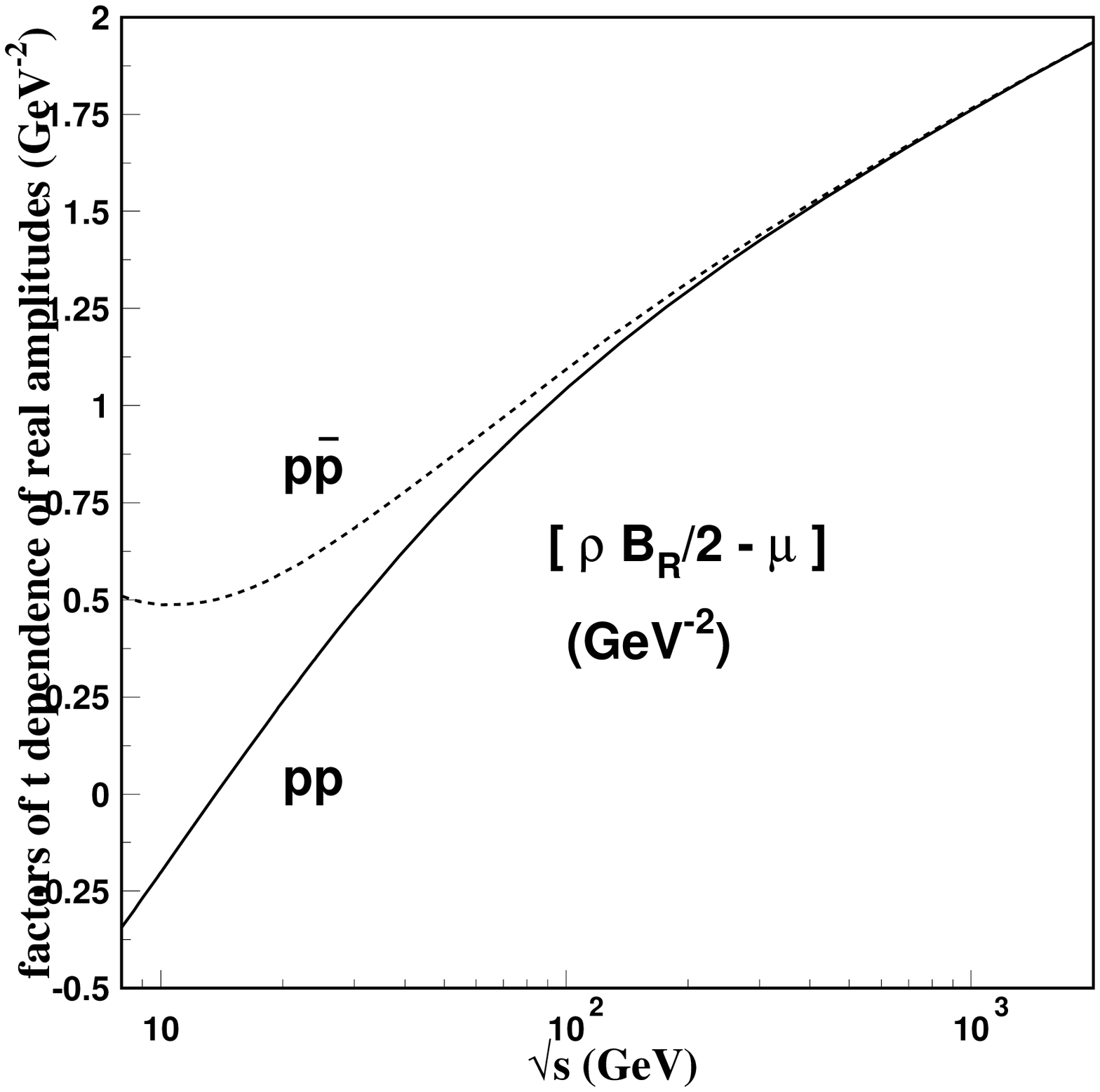}
\caption{ (a) Energy dependence of $\rho({\rm pp})=  {\rm Re}F_{\rm pp}(x,0)/{\rm Im}F_{\rm pp}(x,0)$  and
$\rho({\rm p\bar p})=  {\rm Re}F_{\rm p\bar p}(x,0)/{\rm Im}F_{\rm p\bar p}(x,0)$ predicted by DR, with illustrative values
$0$ and  $-300$ for the
subtraction constant $K$. (b) With $\sigma$ and $B_I$ (for   pp and $\rm p \bar p$ systems) given
as inputs, the combinations of parameters $\rho\, B_R/2 - \mu $  (for pp and ${\rm p \bar p}$ ) are predicted by the
DRS, offering  efficient control in the analysis   of $d\sigma/dt$ data.
 }
\label{amplitudes-fig}
\end{figure*}

The determination of the dimensionless subtraction constant  $ K $, that is particularly
important   at low energies, uses   experimental information
on the real part  of both pp and $\rm p \bar p $ systems.
In the illustrative Fig.~\ref{amplitudes-fig} (a),   values $K=0$ and
$K=-300$ enter   as  examples to show the influence in the predictions of
the real amplitudes for $t=0$ for low energies.

\medskip
\noindent{\bf Structure of the Real Part.}

We can learn more about the  $|t|$  dependence of the real amplitudes through the
investigation of the dispersion relations for slopes.
From studies at high energies \cite{LHC7TeV,X1800,LHC8TeV} it is known that the real
part has a non-trivial structure in the forward range,  presenting a zero that approaches $|t|=0$
 with increasing energy, in agreement with a theorem by A. Martin \cite{Martin}.
Since the pure exponential form cannot have a zero, the real part must be described
by a more sophisticated structure.
Thus we may assume for the real parts  the  forms
 \begin{eqnarray}
&&(1/{2m^2x})~ {\rm Re}\, F_{\rm pp} (x,t)= \sigma_{\rm pp}~[\rho_{\rm pp}-\mu_{\rm pp}\, t]
 ~e^{-B_{R}^{\rm pp}\,|t|/2}~,
\label{real_forward_pp}  \\
 &&(1/{2m^2x})~{\rm Re}\,F_{\rm p\bar{p}} (x,t)= \sigma_{\rm p\bar{p}}~[\rho_{\rm p\bar{p}}-\mu _{\rm p\bar{p}}\, t]
 ~e^{-B_{R}^{{\rm p\bar{p}}}\,|t|/2} ~.
\label{real_forward_ppbar}
\end{eqnarray}
The derivatives with respect to $t$ at the origin give
\begin{eqnarray}
&&(1/{2m^2x})\,\frac{\partial~ {\rm Re}\, F_{\rm pp} (x,t)}{\partial t}\Big|_{|t|=0}
=  \sigma_{\rm pp} \, \left[ \frac{1}{2} \,\rho_{\rm pp}\,B_{R}^{\rm pp} -\mu_{\rm pp} \right] ~,
\label{derivative_pp}  \\
&&(1/{2m^2x}) \,\frac{\partial~ {\rm Re}\, F_{\rm p\bar{p}} (x,t)}{\partial t}\Big|_{|t|=0}
=  \sigma_{\rm p\bar{p}}\, \left[\frac{1}{2} \,\rho_{\rm p\bar{p}}\,B_{R}^{\rm p\bar{p}}- \mu_{\rm p\bar{p}}\right] ~.
\label{derivative_ppbar}
\end{eqnarray}
Terms with  higher  powers do not contribute to the derivative at $t=0$.

We may observe particularly   the combination of parameters
\begin{equation}
\rho \,B_R/2-\mu
\end{equation}
that together with $\rho$  for the pp and  p\=p  systems,
are the predictions of DR and DRS for the description of $d\sigma/dt$ data in the forward range.
 These quantities are shown in Fig. \ref{amplitudes-fig}.
The asymptotic (high energy) values are
$ \rho \approx \pi /\log(x) $   and $  \rho\, B_R/2-\mu \approx \pi  b_2 \log(x)$.
The quantities   $\rho$, $B_R$ and $\mu$,  must be extracted from the
analysis of the $d\sigma/dt$ data and compared with these predictions.

\section {Final Remarks  \label{conclusions} }

The paper presents advances in the formulation and use of dispersion relations in the treatment of
pp and p\=p scattering  in the forward range.
The main points of our results are the reviewed below.

 First, we give the proof of the solution of the general form of  principal value integrals,
arising from terms of form $  x^\lambda \,\log^n{x} $ in the imaginary amplitudes.  The sum
of such terms constitutes  the established representation for the energy dependence of total
cross sections \cite{PDG}, so that the  proof is of fundamental importance for the area,
closing a long period in which the solution of the singular integrals of the dispersion relations
was the main technical difficulty.

The solution is obtained from the  study of a new property found by the present authors
\cite{EF-K-Sesma3}  for the Lerch's transcendent, that is a well defined function of Mathematical Analysis.
In Section \ref{Lerch}  we start from the theorem proving the new integral representation of Lerch
functions and construct the analytic solution for the required integrations.

To investigate   new consequences of the DR principles,
$t$-dependent extensions are written for the imaginary amplitudes,  with   exponential
factors of energy dependent slopes. These extensions are obvious  and universally
adopted in the description of the forward peak of the elastic differential cross
section (the famous diffraction peak). Their use in the framework of dispersion relations
was first introduced \cite{EF2007} in 2007   and shown to be important tool in the
control of parameters of forward elastic scattering.

Parametrizations are introduced for the energy dependence  of the imaginary slopes $B_I^{\rm pp}$,
$B_I^{\rm p \bar p}$, describing well the known data, with analytical structure
formed with  combination of terms of the same basic form $  x^\lambda \,  \log^n{x} $.

Derivatives with respect to $t$ of the DR expressions with the $t$-dependences give origin
to new connections between real and imaginary amplitudes. Taking $t=0$ in these
expressions we obtain the derivatives of the real parts  in terms of PV integrals that
we know how to solve. The novel expressions are here  called Dispersion Relations for Slopes,
DRS. Full expressions are written for these relations  in terms of the given inputs of
the imaginary parts, using the exact PV  solutions.

It is stressed that DR and DRS together form important frame for the analysis of elastic
scattering, totally based on consequences  of the principles of analyticity and causality
that are basis of the theory of dispersion relations.

The relation of these results to the phenomenology of forward elastic scattering is
explored, exhibiting the evaluation  of observable quantities.  Particularly
interesting is the study of the behavior  of the real amplitudes in the forward range,
with identification of constraints that are determined by DR and DRS and point to controls
of the scattering parameters.

\section*{Acknowledgments}

The authors wish to thank the Brazilian agencies CNPq  and CAPES and the Gobierno de Arag\'on (Spain) for financial support.

\appendix  \section{Properties of a Family of Integrals     \label{PV_Integrals_App} }

The purpose of this appendix is to discuss properties of the family of integrals
\begin{equation}
I(\nu,\lambda,x)=\int_{1}^{+\infty}\frac{x^{\prime\lambda}\,(\log x^\prime)^\nu}{x^{\prime 2}-x^2}\,dx^\prime     \label{bati1}
\end{equation}
of parameters $\nu$, $\lambda$, and $x$, which may be complex. The integration takes place along the real axis, from $1$ to $+\infty$. Integrals with a different integration path could be reduced to the form (\ref{bati1}) by adequate changes (translations, rotations, and/or dilatations) of the integration variable $x^\prime$. Along this appendix we assume that
\begin{equation}
{\rm Re}\,\nu >-1\,,  \qquad {\rm Re}\,\lambda < 1\,,      \label{bati2}
\end{equation}
necessary conditions for the finiteness of the integral in the right-hand side of the definition (\ref{bati1}).

First of all, notice that, although $x$ appears as one of the parameters of the integral in (\ref{bati1}), this depends on $x$ through $x^2$. In other words, $I(\nu,\lambda,x)$ and $I(\nu,\lambda,-x)$ are exactly the same. Consequently, the discussion below considers only values of $x^2$. To avoid ambiguities, we assume in what follows that $-\pi/2<\arg(x)\leq\pi/2$.

To start with, let us distinguish three cases, according to the value of $x^2$:
\begin{itemize}
\item (i) $x^2\notin [1,+\infty)$. The integrand in the right-hand side of Eq.~(\ref{bati1})
is in this case regular on the whole integration interval.
\item (ii) $x^2\in (1,+\infty)$. The integrand presents a first order pole at $x^\prime=x$.
We assume in this case that the integral in the left-hand side of Eq.~(\ref{bati1}) is understood as its Cauchy principal value.
\item (iii) $x^2=1$. The integrand may be singular at the lower boundary of the integration interval.
\end{itemize}

\subsection{Case (i): $x^2\notin [1,+\infty)$}

Let us introduce in the right-hand side of (\ref{bati1}) the change of integration variable
\begin{equation}
x^\prime=e^{t/2}\,,   \label{batii1}
\end{equation}
to obtain
\begin{equation}
I(\nu,\lambda,x)=\frac{1}{2^{\nu+1}}\int_{0}^{+\infty}\frac{t^\nu\,e^{-(1-\lambda)t/2}}{1-x^2e^{-t}}\,dt   ~ .   \label{batii2}
\end{equation}
Recalling the integral definition of the Lerch's transcendent \cite[Eq.~25.14.5]{dlmf},
\begin{equation}
\Phi(z,s,a) = \frac{1}{\Gamma(s)}\int_0^{\infty}\frac{t^{s-1}\,e^{-at}}{1-z\,e^{-t}}\,dt \,,  \label{batii3}
\end{equation}
with
\begin{equation}
{\rm Re}\,s>0\,, \qquad {\rm Re}\,a>0\,, \qquad z\in \mathcal{C}\backslash [1,\infty)\,,  \label{batii4}
\end{equation}
we may write
\begin{equation}
I(\nu,\lambda,x)=\frac{\Gamma(\nu+1)}{2^{\nu+1}}\,\Phi\left(x^2,\nu+1,\frac{1-\lambda}{2}\right)\,,
 \qquad {\rm for} \quad  x^2\in \mathcal{C}\backslash [1,\infty)\,.    \label{batii5}
\end{equation}

In the case of $x$ belonging to the open unit disc, that is, $|x|<1$, or if $|x|=1$ and ${\rm Re}\,\nu >0$, the
series representation of the Lerch's transcendent, Eq.~(\ref{i1}),
can be used to obtain for our integral
\begin{equation}
I(\nu,\lambda,x)=\Gamma(\nu+1)\,\sum_{m=0}^\infty\frac{x^{2m}}{(2m+1-\lambda)^{\nu+1}}\,, \quad |x|< 1\,,  \quad{\rm or} \quad |x|=1\,, \; {\rm Re}\,\nu>0\,.  \label{batii8}
\end{equation}
For $x$ out of the unit disc, that is, $|x|>1$, we are able to give a convergent series expansion of our integral only in the particular case of integer $\nu=n=0, 1, 2, \ldots$. Our representation of the Lerch's transcendent for $|z|>1$ and $z\notin (-\infty, 1)\bigcup(1, +\infty)$ \cite[Eq.~(31)]{EF-K-Sesma3},
\begin{equation}
\Phi(z,n,a) = \frac{\pi}{(n-1)!}\left[\frac{\partial^{n-1}}{\partial t^{n-1}}
\Big(z^{t}\,\big({\rm sgn}(\varphi)\,{i}-\cot(\pi t)\big)\Big)\right]_{t=-a} -\,\sum_{m=1}^\infty\frac{z^{-m}}{(a-m)^n}\,, \label{batii9}
\end{equation}
where $\varphi=\arg(\ln z)$ and sgn represents the sign function, allows us to write
\begin{eqnarray}
&& I(n,\lambda,x)  =  \frac{\pi}{2x}\frac{\partial^{n}}{\partial \lambda^{n}}
\bigg(x^{\lambda}\Big({\rm sgn}(\theta)\,{i}+\tan(\pi\lambda/2)\Big)\bigg)   - n! \,\sum_{m=1}^\infty\frac{x^{-2m}}{(1\! -\! \lambda\! -\! 2m)^{n+1}}\,,  \nonumber  \\
&& \hspace{150pt}    \quad  |x|>1, \quad \theta=\arg(\ln x).    \label{batii10}
\end{eqnarray}
This equation seems to be not valid for $\lambda=-(2N+1)$, $N=0, 1, 2, \ldots$, since both terms in the right-hand side become singular. However, the sum of both terms is regular, as it can be checked by taking its limit as $\lambda\to -(2N+1)$.

\subsection{Case (ii): $x^2\in (1,+\infty)$}

We define in this case
\begin{equation}
I(\nu,\lambda,x)= \mathbf{P}\int_{1}^{+\infty}\frac{x^{\prime\lambda}\,(\log x^\prime)^\nu}{x^{\prime 2}-x^2}\,dx^\prime  ~,    \label{batiii1}
\end{equation}
where the symbol $\mathbf{P}$ indicates the Cauchy principal value of the integral. The change of integration variable mentioned in Eq.~(\ref{batii1}) leads to
\begin{equation}
I(\nu,\lambda,x)=\frac{x^{-2}}{2^{\nu+1}}\,\mathbf{P}\int_{0}^{+\infty}\frac{t^\nu\,e^{(1+\lambda)t/2}}{x^{-2}e^{t}-1}\,dt  ~.   \label{batiii2}
\end{equation}
In the case of $\nu$ being a non-negative integer, $\nu =n=0, 1, 2, \ldots$, the right-hand side can be written as a convergent series. In a paper dealing with the Lerch's transcendent, we have found \cite[Eq.~(7)]{EF-K-Sesma3} that, for positive real $z<1$ and complex $a$ such that ${\rm Re}\,a<1$,
\begin{eqnarray}
\mathbf{P}\int_0^{\infty} \frac{t^{n}\, e^{at}}{z\, e^t-1}\,dt =
-\pi\,\frac{\partial^{n}}{\partial a^{n}}\left(z^{-a}\,\cot(\pi a)\right)+(-1)^n\,n!\,\Phi(z,n+1,a)\,,   \label{batiii3}
\end{eqnarray}
which, substituted in (\ref{batiii2}) with $\nu=n$, gives
\begin{eqnarray}
I(n,\lambda,x)=\frac{\pi}{2x}\,\frac{\partial^{n}}{\partial\lambda^{n}}\left(x^{\lambda}\,\tan(\pi\lambda/2)\right)  +\frac{(-1)^n\,n!}{2^{n+1}x^2}\,\Phi\left(\frac{1}{x^2},n+1,\frac{1+\lambda}{2}\right)\,,   \label{batiii4}
\end{eqnarray}
or, by using the series representation of the Lerch's transcendent,
\begin{eqnarray}
I(n,\lambda,x)=\frac{\pi}{2x}\,\frac{\partial^{n}}{\partial\lambda^{n}}\left(x^{\lambda}\,\tan(\pi\lambda/2)\right)  + (-1)^n\,n!\,\sum_{m=1}^\infty\frac{x^{-2m}}{(2m-1+\lambda)^{n+1}}\,.    \label{batiii5}
\end{eqnarray}

\subsection{Case (iii): $x^2=1$}

Now we have
\begin{equation}
I(\nu,\lambda,1)=\int_{1}^{+\infty}\frac{x^{\prime\lambda}\,(\log x^\prime)^\nu}{x^{\prime 2}-1}\,dx^\prime  ~ .     \label{bativ1}
\end{equation}
The change of integration variable (\ref{batii1}) allows one to write
\begin{equation}
I(\nu,\lambda,1)=\frac{1}{2^{\nu+1}}\int_{0}^{+\infty}\frac{t^\nu\,e^{-(1-\lambda)t/2}}{1-e^{-t}}\,dt  ~ .    \label{bativ2}
\end{equation}
For ${\rm Re}\,\nu>0$, bearing in mind the integral representation of the Hurwitz zeta function \cite[Eq.~25.11.25]{dlmf},
\begin{equation}
\zeta(s,a) = \frac{1}{\Gamma(s)}\int_0^{\infty}\frac{t^{s-1}\,{\rm e}^{-at}}{1-e^{-t}}\,dt \,, \qquad {\rm Re}\,s>1\,, \quad {\rm Re}\,a>0\,,\label{bativ3}
\end{equation}
we can write
\begin{equation}
I(\nu,\lambda,1)=\frac{\Gamma(\nu+1)}{2^{\nu+1}}\,\zeta\left(\nu+1,\frac{1-\lambda}{2}\right), \quad {\rm Re}\,\nu >0\,,    \label{bativ4}
\end{equation}
that is, by substituting the series representation \cite[Eq.~25.11.1]{dlmf} of the Hurwitz zeta function,
\begin{equation}
I(\nu,\lambda,1)=\Gamma(\nu+1)\sum_{m=0}^\infty \frac{1}{(2m+1-\lambda)^{\nu+1}}   , \quad {\rm Re}\,\nu >0\, .     \label{bativ5}
\end{equation}
For $-1<{\rm Re}\,\nu \leq 0$, the integrand in the right-hand side of Eq.~(\ref{bativ1}) becomes singular at $x^\prime=1$. A careful analysis is needed in this case.

\end{document}